\documentclass[useAMS,usenatbib]{mn2e}
\usepackage{graphicx}
\newcommand\Msun{M$_\odot$}
\newcommand\Mstar{M$_\star$}
\newcommand\ha{H\,{$\alpha$}}
\newcommand\hb{H\,{$\beta$}}

\newcommand\oii{[O\,{\sevensize II}]}
\newcommand\nii{[N\,{\sevensize II}]}

\newcommand\Aha{$A_{\rm H\alpha}$}
\newcommand\NBH{NB$_{\rm H}$}
\newcommand\apj{ApJ}
\newcommand\apjs{ApJS}
\newcommand\apjl{ApJL}
\newcommand\mnras{MNRAS}
\newcommand\aap{A\&A}
\newcommand\pasj{PASJ}

\title[The nature of faint \oii emitters at $z=1.47$]{
Calibrating [OII] star-formation rates at z$>$1 from dual H$\alpha$-[OII] imaging from HiZELS
} 
\author[M. Hayashi et al.]{%
Masao Hayashi$^{1}$\thanks{E-mail: masao.hayashi@nao.ac.jp},
David Sobral$^{2}$,
Philip N. Best$^{3}$, 
Ian Smail$^{4}$, 
\newauthor
and Tadayuki Kodama$^{1,5}$\\
$^{1}$Optical and Infrared Astronomy Division, National Astronomical Observatory, Mitaka, Tokyo 181-8588, Japan\\
$^{2}$Leiden Observatory, Leiden University, P.O. Box 9513, NL-2300 RA Leiden, THe Netherlands\\
$^{3}$SUPA, Institute for Astronomy, Royal Observatory of Edinburgh, Blackford Hill, Edinburgh, EH9 3HJ, UK\\
$^{4}$Institute for Computational Cosmology, Durham University, South Road, Durham, DH1 3LE, UK\\
$^{5}$Subaru Telescope, National Astronomical Observatory of Japan, 650 North A'ohoku Place, Hilo, HI 96720, USA\\
}

\begin{document}


\pagerange{\pageref{firstpage}--\pageref{lastpage}} \pubyear{2012}

\maketitle

\label{firstpage}

\begin{abstract}
We investigate the relationship between \ha\ and \oii($\lambda3727$)
emission in faint star-forming galaxies at $z=1.47$ with dust
uncorrected star formation rates (SFRs) down to 1.4 \Msun\ yr$^{-1}$,
using data in two narrow-bands from WFCAM/UKIRT and Suprime-Cam/Subaru. 
A stacking analysis allows us to investigate \ha\ emission flux from
bright \oii\ emitters as well as faint ones for which \ha\ is not
individually detected, and to compare them with a large sample of
local galaxies.   
We find that there is a clear, positive correlation between the
average \ha\ and \oii\ luminosities for \oii\ emitters at $z=1.47$,
with its slope being consistent with the local relation. 
\oii\ emitters at $z=1.47$ have lower mean observed ratios of
\ha/\oii\ suggesting a small but systematic offset (at $2.8\sigma$
significance) towards lower values of dust attenuation,
\Aha$\sim0.35$, than local galaxies.      
This confirms that \oii\ selection tends to pick up galaxies which are
significantly less dusty on average than H$\alpha$ selected ones, with
the difference being higher at $z=1.47$ than at $z=0$. The
discrepancy of the observed line ratios between \oii\ emitters at
$z=1.47$ and the local galaxies may in part be due to the samples
having different metallicities.    
However, we demonstrate that metallicity is unlikely to be the main cause.
Therefore, it is important to take into account that the relations
for the dust correction which are derived using \ha\ emitter
samples, and frequently used in many studies of high-$z$ galaxies, may
overestimate the intrinsic SFRs of \oii-selected galaxies, and that
surveys of \oii\ emission galaxies are likely to miss dusty populations.   
\end{abstract}


\section{Introduction}
\label{sec;introduction}

Star formation rate (SFR) is one of the most important properties to
characterise the growth of a galaxy, and the star formation history of the Universe
provides us with a fundamental insight into galaxy evolution.  
Several surveys have been conducted to reveal the star formation
activity in the distant Universe by making use of various indicators
such as ultraviolet (UV) luminosity, nebular emissions such as \ha\
and \oii($\lambda3727$), and infrared (IR) radiation. These have been
used to estimate SFR of galaxies over a wide range of redshifts; 
e.g., UV: \citet{Ouchi2009,Bouwens2011},     
nebular lines: \citet{Shioya2008,Ly2007,Ly2011,Geach2008,Sobral2009,Sobral2012a,Sobral2012b},     
IR: \citet{PerezGonzalez2005,Wardlow2011,Magnelli2009,Magnelli2012,Goto2010}.     
Such surveys have revealed that the redshift range $z=$1--3 is an essential and
intriguing era for the study of galaxy formation and evolution,
since star formation rate density (SFRD) in the Universe gradually
increases toward $z\sim3$ from $z\ga$6, has a peak at $z\sim$1--2, and
decreases sharply from $z\sim1$ toward $z\sim0$
\citep[e.g.,][]{madau1996,Hopkins2006,Sobral2012b}.    
The \ha\ luminosity is widely used to derive the star formation
activity of galaxies at $z\la2$, but other emission-lines at the bluer
end of the galaxy spectral energy distribution, such as the \oii\ line,
are also sufficiently bright to be widely used and can be employed up
to $z\sim4$, after being calibrated relative to \ha. The \ha\ emission 
line is a robust star formation indicator which has been
well-calibrated with data in the local Universe
\citep[e.g.,][]{kennicutt1998}, and is significantly less affected by 
dust extinction than bluer emission lines, such as \oii.    
However, the \ha\ line is redshifted into near-infrared (NIR)
wavelengths for galaxies at $z>0.4$, while the \oii\ line can be observed with
an optical instrument until $z\sim1.7$, and thus many studies/surveys
of star-forming galaxies at $z>0.4$ rely on \oii\ luminosity. Unfortunately, while \oii\
luminosity is in general correlated with the star formation activity,
it is also dependent on the metal abundance and the ionization state
of nebular gas.
The indirect relation with the star formation activity complicates the
estimation of SFR from \oii\ luminosity. Nevertheless, it is
empirically calibrated and extensively utilised as a important SFR
indicator for galaxies at $z\ga1$ \citep[e.g.,][]{kennicutt1998,Moustakas2006,Gilbank2010}.      

In order to better relate the \oii\ luminosity of a galaxy to its SFR,
one can study the ratio of \oii\ to \ha\ luminosities in local
star-forming galaxies.
However, it is not obvious that these calibrations can be blindly
applied for galaxies at higher redshifts. It is therefore important to
investigate their validity at earlier epochs in the Universe by
studying higher redshift galaxies directly.   
According to recent studies, the \oii/\ha\ ratio of galaxies up to
$z\sim1.0$ is on average consistent with that of local galaxies
\citep{Moustakas2006,Weiner2007}. \citet{Sobral2012a} investigated the
\oii/\ha\ ratio for star-forming galaxies at $z=1.47$ by using
wide-field, deep imaging with two narrow-band filters which can catch
\ha\ and \oii\ emission from $z=1.47$ galaxies simultaneously. This
study finds relatively little evolution in the line ratio when
compared to lower redshift. However, the detection of both emission
lines for galaxies at $z\sim1.5$ is limited to relatively luminous
galaxies with \ha\ luminosity larger than $\sim10^{42}$ erg~s$^{-1}$
(i.e., SFR $\ga$ 10 \Msun\ yr$^{-1}$), and little is known about the
line ratio and nature of the more numerous fainter star-forming
galaxies. Fortunately, the \oii\ data obtained with the 
Subaru telescope can reach significantly lower luminosities in \oii\
and, using a stacking analysis, the \oii/\ha\ line ratio can now be
studied down to galaxies with significantly lower SFRs. 

The structure of this paper is as follows. The data used in this paper are
described in \S~\ref{sec;data}. The stacking analysis procedure and
the results from the stacked images are shown in 
\S~\ref{sec;stacking}. In \S~\ref{sec;discussions}, we show the ratios 
of \ha\ to \oii\ luminosities for galaxies at $z=1.47$, and compare
them with those of local galaxies. We then discuss their implications
with respect to the evolution in dust extinction and metallicity of
\oii -selected galaxies at $z=1.47$. Our conclusions are presented in
\S~\ref{sec;conclusions}.      
Throughout this paper, magnitudes are presented in the AB system, and
we adopt cosmological parameters of $h=0.7$, $\Omega_{m}=0.3$ and
$\Omega_{\Lambda}=0.7$.

\section{Observations and selection}
\label{sec;data}

The bulk of the data used in this paper have been obtained as part of
the High Redshift Emission Line Survey (HiZELS) targeting \ha\
\citep{Geach2008,Sobral2009,Best2010}, and as part of a matched \oii\
follow-up survey \citep{Sobral2012a,Sobral2012b}. The details of the
data are described in two papers by \citet{Sobral2012a,Sobral2012b},
and therefore we only summarise them briefly here. 

\begin{figure}
 \begin{center}
 \includegraphics[width=\linewidth]{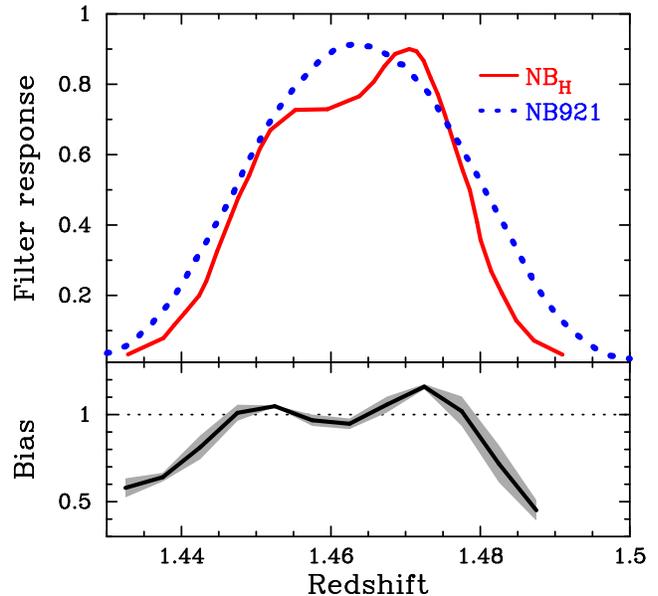}
 \end{center}
 \caption{ The upper panel shows the response curves of the \NBH\ (red
 solid line) and NB921 (blue dotted line) narrow-band filters as a
 function of redshift which corresponds to the wavelengths where \ha\
 and \oii\ lines are detectable, respectively. 
 The lower panel shows the bias of observed line ratio of \ha/\oii\ as
 a function of redshift, which is caused by the slight differences of filter
 profiles. The grey region shows the 1$\sigma$ dispersion. Since
 the profile of the NB921 filter is slightly broader, the \ha/\oii\ ratio
 can be underestimated for galaxies at the edge of the redshift range where
 the NB filters can catch the emission lines. However, since most of
 the galaxies selected with the NB filters are expected to be located at
 redshifts near the peak of filter profile, it is likely that the
 difference of filter profiles, which we correct for by 3\%, does not
 significantly influence the measurement of the line ratio.
  }   
 \label{fig;filters}
\end{figure}

\subsection{\ha\ and \oii\ emitters at $z=1.47$}
\label{sec;sample}

We use two narrow-band imaging datasets to obtain our samples of
galaxies at $z=1.47$ in two distinct square degree areas: the UKIDSS Ultra Deep
Survey \citep[UDS:][]{Lawrence2007} and the Cosmological Evolution
Survey \citep[COSMOS:][]{Koekemoer2007,Scoville2007} fields.
\NBH\ narrow-band data ($\lambda_c=1.617$\micron\ and
$\Delta\lambda=0.021$\micron) were taken with the wide-field camera
\citep[WFCAM;][]{Casali2007} on UKIRT, while NB921
narrow-band data ($\lambda_c=9196$\AA\ and $\Delta\lambda=132$\AA)
were taken with Subaru prime focus camera
\citep[Suprime-Cam;][]{Miyazaki2002} on the Subaru Telescope
(Figure~\ref{fig;filters}). Combining with broad-band imaging data in
$H$ and $z'$ which cover the same wavelength range as the individual
narrow-bands, to estimate the continuum level of the spectrum underlying
the emission line, the narrow-band imaging can measure \ha\ and \oii\
emission, simultaneously, for galaxies at $z=1.47$ \citep{Sobral2012a,Sobral2012b}. 

The \NBH\ and $H$ images in the UDS field cover an
effective area of 0.78 deg$^2$ where the regions with bad quality
caused by cross-talk and bright stars are masked, while the images in
the COSMOS field have an effective area of 1.6 deg$^2$.
The 5$\sigma$ limiting magnitudes in \NBH\ are $\sim$22.1 and
$\sim$21.9 in the UDS and COSMOS fields, respectively, although the
depth is slightly dependent on the position (see Figure 2 and Table 2
in \citet{Sobral2012b} for the details).  

The NB921 data in UDS field are drawn from the archive of
Suprime-Cam \citep{Ouchi2010} and reduced as described in \citet{Sobral2012a},
while public data are available for $z'$ \citep{Furusawa2008}.
In the COSMOS field, the NB921 data were taken in service mode with
the Subaru telescope in December 2010 \citep{Sobral2012b}, and they
cover 69\% of the region where the \NBH\ and $H$ data are available
(i.e., 1.1 deg$^2$). The 5$\sigma$ limiting magnitudes in NB921 are
25.8 and 24.0 in the UDS and COSMOS fields, respectively.  

Catalogues of \ha\ emitters at $z=1.47$ in both the UDS and COSMOS
fields are presented in \citet{Sobral2012b}. They include 188 \ha\
emitters in the UDS down to an \ha\ flux of $4.5\times10^{-17}$
erg~s$^{-1}$~cm$^{-2}$ (which corresponds to an \ha\ luminosity
$\log$(L$_{{\rm H}\alpha}$/erg~s$^{-1}$)=41.78 if the galaxies are at
$z=1.47$) and 325 \ha\ emitters down to $2.7\times10^{-17}$
erg~s$^{-1}$~cm$^{-2}$ ($\log$(L$_{{\rm H}\alpha}$/erg~s$^{-1}$)=41.56) 
in the COSMOS field, respectively.      
The number densities of the \ha\ emitters are 6.7 and
8.2$\times10^{-2}$ arcmin$^{-2}$ above the flux limits in UDS and
COSMOS fields, respectively.  

The catalogues of \oii\ emitters at $z=1.47$ are updated from those of
\citet{Sobral2012a}. Emission line galaxies are selected as galaxies
with colour excess larger than 5$\sigma$ (i.e., $\Sigma>5$) and
equivalent width (EW) larger than 25\AA, which corresponds to
EW$_0>10$\AA\ in rest frame at $z=1.47$, where photometry is conducted
with 2\arcsec\ diameter aperture. Then, \oii\ emitters at $z=1.47$ are
identified based on the colours, photometric redshift, and
spectroscopic redshift \citep[c.f.][]{Sobral2012b}.  We note that the
spectroscopic redshift information indicates that a significant number
of galaxies at $z\sim1.3$, selected as narrow-band emitters due to
strong Balmer/4000\AA\ break, are incorrectly included as
\oii\ emitters. However, most of these contaminants are removed from
the sample by filtering out galaxies with EW$_0<20$\AA\ $\cap$
$i-z>0.55$.  Thus, the numbers of \oii\ emitters selected are 2735
down to $4.2\times10^{-18}$ erg~s$^{-1}$~cm$^{-2}$ 
($\log$(L$_{{\rm [O II]}}$/erg~s$^{-1}$)=40.76) in the UDS field, and
718 down to $1.3\times10^{-17}$ erg~s$^{-1}$~cm$^{-2}$
($\log$(L$_{{\rm [O II]}}$/erg~s$^{-1}$)=41.24) in the COSMOS field,
respectively. The number densities of the \oii\ emitters are 0.97 and
0.18 arcmin$^{-2}$ above the flux limits in the UDS and COSMOS fields,
respectively.  

\begin{figure}
\begin{center}
  \includegraphics[width=\linewidth]{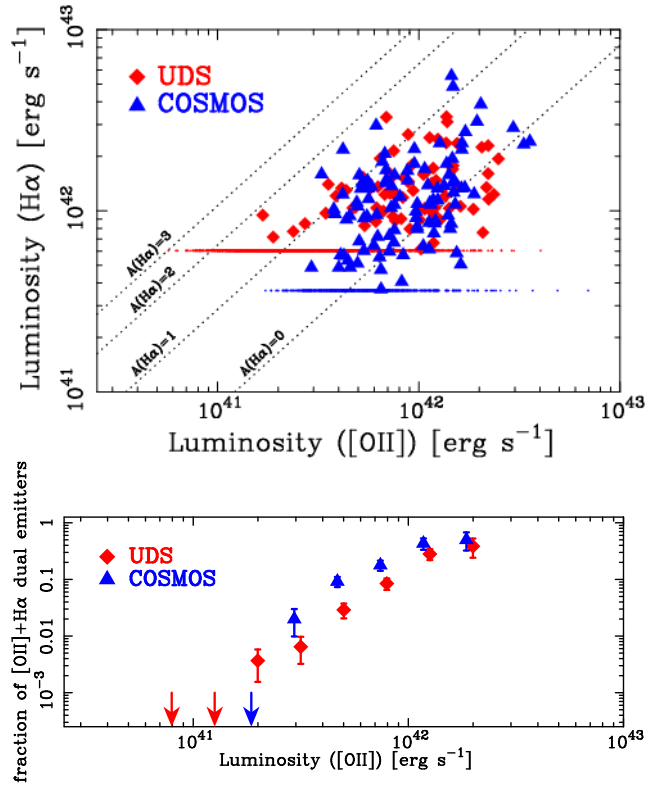} 
  \includegraphics[width=\linewidth]{fig2b.eps} 
\end{center}
 \caption{ (Upper panel) The variation of \ha\ luminosity as a
  function of \oii\ luminosity for the \oii+\ha\ dual emitters at
  $z=1.47$, including the upper limits of the \ha\ luminosity for the
  \oii\ emitters without a detection of \ha\ emission. 
  Red symbols show the \oii\ emitters in the UDS field, and blue
  ones show those in the COSMOS field. The lines show the ratio of
  \ha\ to \oii\ for different levels of dust attenuation \citep{Sobral2012a}.
  (Lower panel) The fraction of \oii\ emitters with \ha\ emission
  detected as a function of \oii\ luminosity. 
  Note that the COSMOS fractions are typically higher since the \NBH\
  data are $\approx 0.2$\,dex deeper in that field. The arrows
  indicate an upper limit to the fraction when there is no dual
  emitter in the luminosity bin. Errors in the fraction are estimated
  based on Poissonian statistics. Note that the dual emitters with
  detections in both \oii\ and \ha\ are mainly limited to galaxies
  with \oii\ luminosities higher than $\sim10^{42}$ erg~s$^{-1}$.
}
 \label{fig;catalogue}
\end{figure}

By matching the \oii\ emitters with the \ha\ emitters, Figure
\ref{fig;catalogue} shows the relation between \oii\ and \ha\
luminosities for the \oii\ emitters in both the UDS and COSMOS fields,
and the fraction of \oii\ emitters with a significant \ha\ emission
detection as a function of \oii\ luminosity. As described in
\S\ref{sec;introduction}, the dual emitters with detections in both
\oii\ and \ha\ are mainly limited to galaxies with \oii\ luminosities
higher than $\sim10^{42}$ erg~s$^{-1}$. For \oii\ emitters with \oii\
luminosities lower than $\sim10^{42}$ erg~s$^{-1}$, the fraction is
less than $\sim10\%$, clearly showing that we can only investigate the
relation between \oii\ and \ha\ luminosities individually for a small
fraction of faint \oii\ emitters.    

\subsection{Differences between the profile of the two narrow-band filters}
\label{sec;NBprofiles}

Although the \NBH\ and NB921 filters are well matched and enable us to
detect both \ha\ and \oii\ emission lines from individual star-forming
galaxies at $z=1.47$, there is a slight difference in the redshift
coverage (Figure~\ref{fig;filters}). The redshift coverage for \oii\
lines detected by NB921 filter is a slightly wider than that by \NBH\
filter. It is possible that this difference causes us to underestimate
the \ha\ luminosity for \oii\ emitters at redshifts corresponding to the
edge of \NBH\ filter where the transmission is lower than that of
NB921. To quantitatively evaluate how accurately the \ha/\oii\ line ratio
can be measured by the two narrow-band images and how this effect may 
influence the results of the stacking analysis, a simulation was
conducted following \citet{Sobral2012a} (see \S 4.5 of their paper
for more details). In summary, we make a sample of galaxies at
$z=$1.40--1.52 with intrinsic line ratios between 0 and 2.0, where the
distribution of \oii\ luminosity that galaxies have is based on the
\oii\ luminosity function, and then investigate the measured \ha/\oii\
line ratios with the two NB filters. Figure~\ref{fig;filters} also
shows the result of the simulation: the bias (i.e., the difference
between a measured \ha/\oii\ and an input \ha/\oii\ ratio) is
distributed around unity in most of the range of the filter profile,
although in the edge of the profile the observed line ratio is biased
towards lower values. However, we make sure that the average value of
bias is fairly close to unity. Because we focus on the average ratio
of \ha/\oii\ by a stacking analysis in this paper, the slight
difference of filter profile has only a small effect on the results  
over the whole investigated range of \oii\ luminosity.   
To further quantify this, we investigate the bias values for galaxies
with a given range of {\it observed} \oii\ luminosity similar to
luminosity bins shown in Table \ref{tbl;OII}; we find that the \oii
-selected galaxies have a redshift distribution peaking at $z\sim1.46$
and with most sources having redshifts where the filter response is
high, leading to an average value of the bias of $\sim$0.97 in all
bins, namely close to unity again. Thus, although the \ha/\oii\ ratio
can be underestimated by 3\% (which we correct for), we conclude that
the difference of the profiles in the two NB filters gives no
significant influence on the results we find in this paper.

\subsection{Stellar mass for emitters at $z=1.47$}
\label{sec;stellarmass}

Stellar masses are estimated by an SED-fitting method for \ha\ and \oii\
emitters at $z=1.47$ in both fields consistently (Sobral et al. in
preparation). The details of the procedure in the SED-fitting are described
in \citet{Sobral2011}. The SED templates are created with the stellar
population synthesis model by \citet{BC03} and \citet{Bruzual2007}
under the assumption of a \citet{Chabrier2003} initial mass function
(IMF) and a range of exponentially declining star formation histories. 
In addition, the dust extinction law of \citet{calzetti2000} is used
in the SED-fitting. 
Note that we have confirmed that the mass estimates are robust against
the contribution of emission lines to the broad-band photometry, as
they are based to fits to $\sim30$ photometric bands and hence line
contamination in one or two bands does not have a strong influence. 
Among \oii\ emitters, stellar masses are derived
for 2708 (99.0\%\ of the sample) and 706 (98.3\%) \oii\ emitters in
the UDS and COSMOS field, respectively. The others, i.e., $\sim1\%$ of
the samples, do not have reliable stellar mass due to large errors in
photometry and/or non-detection in most bands, suggesting that such \oii\
emitters are likely to be less massive galaxies.   
Galaxies for which the SED fitting fails are found to be very faint in
the rest-frame optical (i.e., $K$-band), which supports the indication
that they are likely to be low-mass galaxies. 
In this paper, we restrict analysis to \oii\ emitters with stellar
masses larger than $\log$(\Mstar/\Msun)=9.5. This mass cut is made in
order to maximise the number of \oii\ emitters included in the sample,
but without picking up many lower mass galaxies for which the survey
becomes incomplete. In addition, the SDSS sample described below,
which is used for the comparison of galaxies at $z=1.47$ with local
galaxies, is also highly complete above this mass limit. After applying the
mass cut, the numbers of \oii\ emitters are 643 and 212 in the UDS and
COSMOS fields, respectively.

\subsection{A comparison sample of galaxies at $z\sim0.1$}
\label{sec;sdsssample}

\begin{figure}
 \begin{center}
 \includegraphics[width=\linewidth]{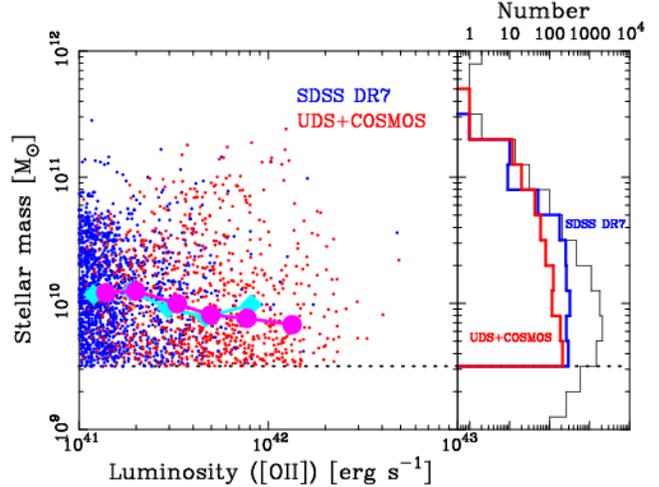}\\
 \end{center}
 \caption{
 The left panel shows stellar mass as a function of \oii\ luminosity. Red
 dots show \oii\ emitters at $z=1.47$ in the combined sample and blue
 dots show galaxies in the SDSS sample with similar median mass
 distribution to that of \oii\ emitters at $z=1.47$. Magenta circles
 and cyan diamonds show the median stellar mass of galaxies in each \oii\
 luminosity bin for the samples of \oii\ emitters and SDSS galaxies, 
 respectively. The right panel shows mass distribution of each
 samples: the black thin histogram shows the whole sample of SDSS
 galaxies with $\log$(L$_{{\rm [O II]}}$/erg~s$^{-1}$)$>$41 and EW$_0>10$\AA\ at
 $z\sim0.1$, the blue thick one shows those of local galaxies extracted from the 
 SDSS sample, and the red thick histogram shows the mass distribution of \oii\ emitters
 at $z=1.47$ in the combined sample including the UDS and COSMOS fields.   
}
 \label{fig;mass_dist}
\end{figure}

We also use SDSS DR7 data to compare the results obtained for \oii\
emitters at $z=1.47$ with local galaxies \citep{Abazajian2009}. 
Galaxies in the redshift range of $z=$0.07--0.1 are extracted from the
SDSS spectroscopic catalogue.      
This redshift range is chosen so that the galaxies have small enough
apparent angular sizes that most of their light is included in the
fibre whilst guaranteeing that the sensitivity is still very high. An
aperture correction on the emission line luminosities is still
required, since the spectroscopic measurements are done with
3\arcsec-diameter fibres. The fractional flux loss from the fibre is
estimated from the ratio of total mass to fibre mass according to the
same procedure adopted in \citet{Sobral2012a}.  
Moreover, note that the SDSS spectrophotometric calibration takes
account of any wavelength dependence of the seeing, so the emission
line flux ratios are also unaffected by this \citep{Adelman-McCarthy2008}.

By applying our selection criteria, we have selected
8285 SDSS galaxies with $\log$(L$_{{\rm [O II]}}$/erg~s$^{-1}$)$>$41 and
EW$_0>10$\AA. Among these, only galaxies with detected  \ha\ emission are
used. However, we note that all but 38 (0.46\%) galaxies have \ha\ line
detected, implying that the removal of galaxies without \ha\ detection
gives no significant bias in the study of \ha/\oii\ line ratios.     
We are aware that the SDSS spectroscopic sample is magnitude-limited,
i.e.~roughly mass-limited. However, as described in
\S\ref{sec;stellarmass}, the criterion of $\log$(\Mstar/\Msun)$>$9.5 
is applied to both the SDSS sample and the \oii\ emitter samples at
$z=1.47$ so that high completeness is kept above the mass limit for
both samples; this was another factor driving our choice of redshift range
for the SDSS sample. After applying our mass limit we obtain our SDSS
sample, which contains 7271 galaxies.   
  
The SDSS sample is used to compare the \ha/\oii\ ratios of
\oii\ emitters at $z=1.47$ with those of local galaxies at
$z$=0.07--0.1 in \S\S \ref{sec;stacking} and \ref{sec;discussions}. We 
therefore must assure that both samples are fully comparable and that
differences between them are not arising from a different distribution
in e.g. stellar mass. Since correlations between SFR and stellar mass
have been claimed at $z\sim2$ \citep[e.g.,][]{Daddi2007}, it is
possible that \oii\ luminosity as well as \ha\ luminosity and metallicity are
also correlated with stellar mass for the $z=1.47$ \oii\ emitter
samples. We thus use the SDSS sample to construct a sample of local
galaxies with a similar distribution of mass to that of \oii\
emitters at $z=1.47$ in order to reduce any mass-dependent
bias on this study and so allow a robust comparison.        
Galaxies in the SDSS sample are selected at random so that the mass
distribution is the same as that of \oii\ emitters at $z=1.47$ in each
bin of \oii\ luminosity as shown by Figure \ref{fig;mass_dist}. The
SDSS sample tends to include galaxies with fainter \oii\ luminosity,
while \oii\ emitters at $z=1.47$ have brighter luminosity. Since the mass
distribution is dependent on the \oii\ luminosity, the histograms of
stellar mass for all galaxies in each sample are not in perfect 
agreement. However, the median masses in each luminosity bin are in good
agreement (Figure \ref{fig;mass_dist}). In the following sections, we
use the matched SDSS sample containing 1656 galaxies at
$z$=0.07--0.1 to compare with the results for \oii\ emitters at
$z=1.47$. 

\section{Analysis and results}
\label{sec;stacking}

\begin{figure}
 \begin{center}
 \includegraphics[width=1.0\linewidth]{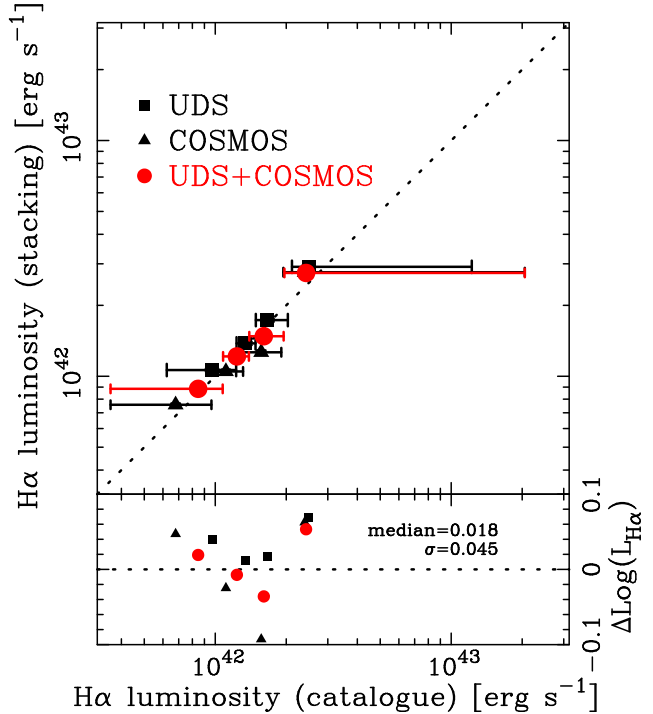}
 \end{center}
 \caption{Comparison between \ha\ luminosities measured on the stacked
 (\NBH--$H$) image and median values of \ha\ luminosities extracted
 from the catalogues, for \ha\ emitters which are individually
 detected as galaxies with ($H$--\NBH) colour excess.  
 We plot both the individual fields and our combined sample.
 The median and standard deviation of $\Delta\log$L$_{\rm H\alpha}$,
 the difference between both \ha\ luminosities, are 0.018 and 0.045,
 implying good agreement between the two \ha\ luminosities, which
 confirms the validity of our stacking analysis procedure.   
 }   
 \label{fig;haha}
\end{figure}

In order to investigate the relation between \ha\ and \oii\
luminosities for galaxies at $z=1.47$, we stack and average the \NBH\
images from which the corresponding $H$ image is subtracted (clipping
the pixels which deviate more than 3$\sigma$), and then estimated the
average \ha\ flux for \oii\ emitters at $z=1.47$ in the UDS and COSMOS
fields. Note the results do not change if a median stacking is carried
out instead.  

\subsection{Stacking method}
\label{sec;method}

For all the samples of galaxies to be stacked, a 80\arcsec$\times$80\arcsec\
region around individual galaxies is extracted from both \NBH\ and $H$
images. Some galaxies are rejected from each sample, because they are found
in masked regions in the \NBH\ image, where some artefacts and regions
with low quality caused by bright/saturated stars are seen.  
Then, the zero point of magnitude and the point spread
function (PSF) are matched between the \NBH\ and $H$ images for individual
galaxies. The original \NBH\ and $H$ images are scaled to have same
zero point of magnitude. We note that \citet{Sobral2012b} show that
small colour-corrections in ($H$--\NBH) colour are required to trace the
underlying continuum level and estimate emission line flux
correctly, simply because the \NBH\ filter is not in the centre of the
$H$-band. We thus performed the correction using the ($J-H$) colour
according to \citet{Sobral2012b}. The PSF in the original \NBH\ image
has a FWHM of $\sim0.8$\arcsec, while that in the $H$ image has FWHM
of $\sim1.0$\arcsec. The \NBH\ PSF is therefore degraded to match the
$H$-band seeing by the convolution of a Gaussian kernel. After
performing these corrections for the zero point of magnitude and PSF,
(\NBH--$H$) images are created by the subtraction of the $H$ image
from the \NBH\ image.        

The individual (\NBH--$H$) images are averaged together to make the
stacked (\NBH--$H$) image for each sample. The photometry is
calculated at the centre of the stacked image with various apertures
ranging from 2.0\arcsec\ up to 6.0\arcsec\ diameter. We note that the
photometry with 4.8\arcsec\ diameter aperture is found to recover the
total flux of the emission line on the stacked image, and then the
(\NBH--$H$) colour is converted into an emission line flux. Tuned
aperture sizes could in principle be dependent on the emission line flux,
but we do not find the significant dependence between the size and the
luminosity. Thus, we use a common aperture with 4.8\arcsec\ diameter
to measure the total fluxes in all samples. We also note that both
\ha\ and \nii\ emission lines can contribute to this emission flux,
but no information of \nii\ flux for individual emitters is
available. Therefore, we corrected for the contribution of \nii\ flux
by assuming that \nii/\ha\ ratio of the emitters is 0.22
\citep{Sobral2012a}.

The \nii/\ha\ ratio is known as a indicator of
metallicity, and thus it is probably dependent on the stellar
mass. Indeed, we find for the SDSS galaxies that the \nii/\ha\
line ratio correlates with stellar mass more strongly than with \ha\ or
\oii\ luminosities. Our samples include the emitters with a wide range
of stellar mass, but the median stellar mass in each sample only varies 
by a factor of four, from $5\times10^{9}$\Msun\ to $2\times10^{10}$\Msun. 
According to the mass--metallicity relation that \citet{Yabe2012} have
found for star-forming galaxies at $z\sim1.4$, these median stellar masses
correspond to the \nii/\ha\ ratios of 0.14--0.25, which are comparable
to or slightly lower than the value we assumed. Even if the lowest ratio of
\nii/\ha=0.14 is used instead, the \ha\ luminosities would be increased by
only a factor of 1.07. It should be noted that the assumption of constant
typical value of \nii/\ha\ is reasonable, since average \ha\
luminosities are derived in this study (see also Figure \ref{fig;haha}). 

Contamination in the \oii\ emitter samples would lead to the \ha\
fluxes measured on the stacked images being underestimated, since the
images without \ha\ flux for the contaminants are also stacked.  
Although the bulk of the contaminants are removed using spectroscopic
and photometric redshifts and colour cuts (\S \ref{sec;sample}), some
contamination will remain. 
Using the spectroscopically confirmed NB921 emitters, and accounting
as best as possible for biases in the selection of spectroscopic
targets, the residual rate of the contamination is estimated to be
around 15$\pm$7\%. 
More specifically, it is found that the $z=1.47$ \oii\ emitter sample
could include $\sim5$\% contaminations at $z<1.0$ and $\sim10$\% ones
at $z>1.0$ (most of which are galaxies with a strong Balmer/4000\AA\
break at $z\sim1.3$). 
We have also found that there is no clear trend of the
contamination rate with emission line flux.
Thus, a correction for our best-estimate of the contamination, 15\%,
is applied to the stacked fluxes for the \oii\ emitter samples in an
attempt to account for this.    

AGNs may be included in our emitter samples as contaminants.
However, galaxies hosting a strong type-1 AGN, with their SED
dominated by AGN light, are likely to fail in our SED fitting
procedure because our code does not include an AGN template SED;
they will thus be excluded from our analysis as only galaxies
well-fitted by galaxy SED templates are used. Galaxies hosting an
obscured type-2 AGN may be included as they can be fitted with
galaxy SED templates.  However, such type-2 AGNs usually have a
\ha/\oii\ ratio above unity \citep[e.g.,][]{CidFernandes2010}, and so
any AGN contamination in the z=1.47 \oii\ samples would, if anything,
bias the stacked \ha\ flux upwards, making the underlying trend found
in Section \ref{sec;stack} even stronger.

To estimate the error in the flux measured on the stacked image, 
10,000 identical apertures are distributed at random across blank sky
region around the galaxy, and then the 1$\sigma$ error is derived by
fitting a Gaussian profile to the histogram of the sky counts.

\subsection{Stacking analysis for the emitter samples}
\label{sec;stack}

\begin{figure}
\begin{center}
  \includegraphics[width=1.0\linewidth]{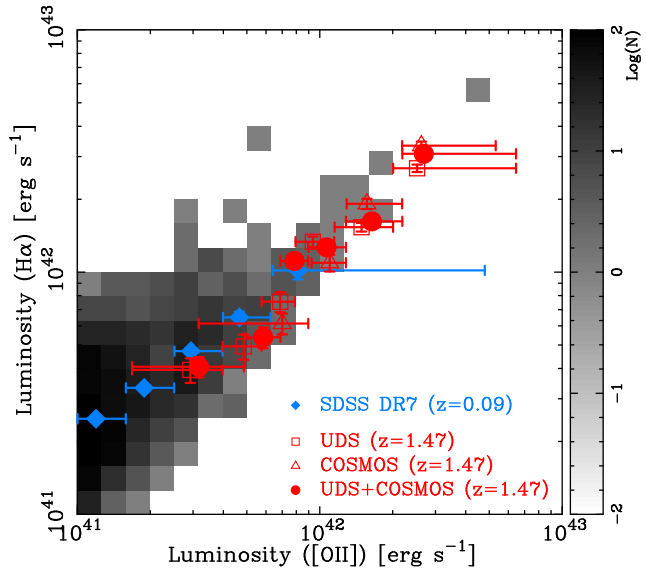} 
\end{center}
 \caption{The comparison between \ha\ and \oii\
  luminosities for \oii\ emitters at $z=1.47$. We show the mean \ha\
  luminosities measured on the stacked (\NBH--$H$) images as a
  function of median \oii\ luminosities in the UDS and COSMOS
  fields. We also plot the combined sample. The gray scale map shows
  the distribution of SDSS galaxies at $z$=0.07--0.1, and blue filled
  circles are median \ha\ luminosities for the local galaxies in each
  \oii\ luminosity bin. In this plot, the \ha\ luminosities are
  corrected for all effects discussed in \S\S \ref{sec;NBprofiles} and
  \ref{sec;method}.  
}
 \label{fig;oiiha}
\end{figure}

\begin{table}
 \caption{The average \oii\ and \ha\ luminosities for \oii\ emitters
 in the combined sample in the UDS and COSMOS fields. The first
 and second columns show the median \oii\ luminosity and the number of
 \oii\ emitters to be stacked in each sample. The third and fourth
 columns show the  mean \ha\ luminosities measured on the stacked
 (\NBH--$H$) images  and the flux ratios of \ha\ to \oii. The
 contribution of \nii\ to the flux measured on the stacked image is
 removed under the assumption that the typical \nii/\ha\ ratio is 0.22
 \citep{Sobral2012a}. 
 The flux ratios are corrected for any effects discussed in \S\S
 \ref{sec;NBprofiles} and \ref{sec;method}.  
 The \oii\ luminosities which are originally measured with a 2\arcsec\
 aperture are corrected by a factor of 1.6 to convert to the total
 luminosities (\S \ref{sec;stack}).}
\begin{center}
\begin{tabular}{cccc}
\hline\hline
 \oii\ luminosity & Number & \ha\ luminosity & flux ratio \\
        (erg s$^{-1}$)  &  &  (erg s$^{-1}$)  &   (\ha/\oii)        \\
\hline
42.43$_{-0.09}^{+0.38}$ &  61 & 42.40$_{-0.01}^{+0.01}$ & 1.15$\pm$0.35 \\ 
42.22$_{-0.11}^{+0.13}$ & 127 & 42.13$_{-0.01}^{+0.01}$ & 0.99$\pm$0.16 \\ 
42.03$_{-0.07}^{+0.08}$ & 125 & 42.02$_{-0.02}^{+0.02}$ & 1.19$\pm$0.13 \\ 
41.90$_{-0.06}^{+0.06}$ & 126 & 41.96$_{-0.02}^{+0.02}$ & 1.41$\pm$0.13 \\ 
41.77$_{-0.08}^{+0.07}$ & 124 & 41.65$_{-0.05}^{+0.04}$ & 0.92$\pm$0.13 \\ 
41.50$_{-0.28}^{+0.19}$ & 246 & 41.53$_{-0.04}^{+0.04}$ & 1.28$\pm$0.37 \\ 
\hline\hline
\label{tbl;OII}
\end{tabular}
\end{center}
\end{table}

Samples of galaxies to be stacked are created from the \ha\ or \oii\
emitters in the UDS and COSMOS fields by dividing them on the basis of
their emission line luminosities, so that they contain nearly
equal numbers of galaxies. However, the number of galaxies for the sample
in the brightest (faintest) luminosity bin is smaller (larger) than the
other samples.  
The stacking analysis was conducted firstly for \ha\ emitter
samples. The analysis for galaxies with \ha\ emission detected 
individually enables us to check our stacking analysis
by comparing our results with individually measured
\ha\ luminosities. 
Figure \ref{fig;haha} shows the comparison between median \ha\
luminosity drawn from the catalogues for each \ha\ emitter sample and
that measured on the stacked image. The median and standard deviation
of $\Delta\log$(L$_{\rm H\alpha}$/erg~s$^{-1}$), the difference
between both \ha\ luminosities, are 0.018 and 0.045, respectively. The
good agreement shows that our procedure in the stacking analysis
recovers well the average of actual individual measurements. We then
moved on the stacking analysis for samples of \oii\ emitters which are
classified on the basis of their \oii\ luminosities.  

Figure \ref{fig;oiiha} shows the average \ha\ luminosities of galaxies
at $z=1.47$ in the individual fields of UDS, COSMOS and the
combination of the two, as a function of their \oii\ luminosity (see
also Table \ref{tbl;OII}). The \ha\ luminosity is estimated from the
stacked \NBH\ -- $H$ image with a 4.8\arcsec aperture (\S
\ref{sec;method}), and the \oii\ luminosity is a median value of
individual \oii\ luminosities of emitters in the sample which are
derived from NB921 imaging. It should be noted that we apply an
aperture correction to the \oii\ luminosities in order to ensure a
matched comparison with the \ha\ luminosities. We obtain this
correction by comparing the  2\arcsec\ photometry
\citep{Sobral2012a,Sobral2012b} and that obtained with 4.8\arcsec
apertures on the stacked NB921--$z'$ images. We find that a correction
of 1.6 is required to recover total stacked luminosities.  
Also, the \ha\ luminosities are corrected by 3\% for the bias caused
by the difference of profile of the two narrow-band filters (\S
\ref{sec;NBprofiles}) and by 15\% for the possible contamination of
emitters at different redshifts that might be included in the sample
(\S \ref{sec;method}).  

There is a clear positive correlation between the
mean \oii\ and \ha\ luminosities, with almost constant luminosity ratio
for \oii\ emitters at $z=1.47$. This fact suggests that the \oii\
luminosity can be easily calibrated as a SFR indicator for galaxies with SFRs
down to 1.4 \Msun yr$^{-1}$, even at $z\sim1.5$.   
The distribution of local galaxies extracted from the SDSS sample is
also shown in the figures, and the median \ha\ luminosities are
plotted in each \oii\ luminosity bin. 
For $z=1.47$ \oii\ emitters with a given \oii\ luminosity, \ha\ luminosities
are on average lower than those of local galaxies, implying that the
ratio of \ha\ to \oii\ is different between local galaxies and \oii\
emitters at $z=1.47$.
To evaluate the statistical significance of the difference in
\ha/\oii\ ratio, we compare the \ha/\oii\ ratios of \oii\ emitters at
$z=1.47$ with those of local galaxies, calculate how significantly the
\oii\ emitters at $z=1.47$ deviate towards a lower value from the
local galaxies in each luminosity bin, and then combine these in
quadrature (using equal weighting) to give a total offset
significance. We find that there is an overall difference in the line
ratio which is significant at the 2.8$\sigma$ level.

The systematic offset is intriguing. It is unlikely that the offset is
caused by a difference in stellar mass, since the mass distributions
of the two sample are matched to each other (\S\ref{sec;sdsssample}). 
Possible systematic effects that might drive the offset are discussed
in the next sub-section, but are unlikely to be able to account for
all of the difference, suggesting a small but genuine cosmic evolution
of the mean \ha\ to \oii\ ratio.

\subsection{Possibility of underestimation of \ha\ luminosity}
\label{sec;underestimation}

One possible cause of bias towards low \ha\ luminosities for \oii\
emitters at $z=1.47$ with a given \oii\ luminosity is contamination in 
the \oii\ emitter samples from emitters at different redshifts which
are not removed by our spectroscopic redshift, photometric redshift and
colour-colour selection cuts. 
As discussed in \S \ref{sec;method}, a correction for the
contamination, 15\%, has already been applied to the stacked fluxes.
If the upper limit to the contamination of 22\%\ is the true
contamination, then the calculated \ha\ luminosities for the \oii\
emitters can be also underestimated by $\sim$7\%.

Another possibility is that the contribution of \nii\ emission is lower
than what we assume in this paper as discussed in \S \ref{sec;method}.
This effect can result in the underestimation of \ha\ luminosity by
$\sim$7\% at most. Finally, there is a bias caused by the difference of
profile of the two narrow-band filters (\S \ref{sec;NBprofiles}) which can
lead to the \ha\ luminosities being underestimated in the wings of the
filter profile. Our simulations have led us to apply a $\sim$3\%
correction factor for this, but these simulations assume a uniform redshift
distribution for the \oii\ emitters across the redshifts probed. If
large-scale structure causes the \oii\ emitters to be clustered at the low
or high redshift end of the filter coverage, the true correction factor
could potentially be a few percent higher.

In order for the \ha\ luminosities at $z=1.47$ to be comparable to
those of galaxies at $z\sim0.1$ for a given \oii\ luminosity, all of
the effects that might potentially cause an underestimation of
\ha\ luminosities would have to be at or above their maximal combined
values of $\sim 15$\%.  Thus, it is likely that the offset of
\ha\ luminosities for galaxies with a given \oii\ luminosity is real,
and the \ha/\oii\ ratios of \oii\ emitters at $z=1.47$ seem to be
shifted towards lower values than those of the local galaxies to some
extent. In the next section, we discuss the implication of the
different \ha/\oii\ ratio in the properties of galaxies.

\section{Discussion}
\label{sec;discussions}

\subsection{Line ratios for the local galaxies}

\begin{figure}
\begin{center}
  \includegraphics[width=1.0\linewidth]{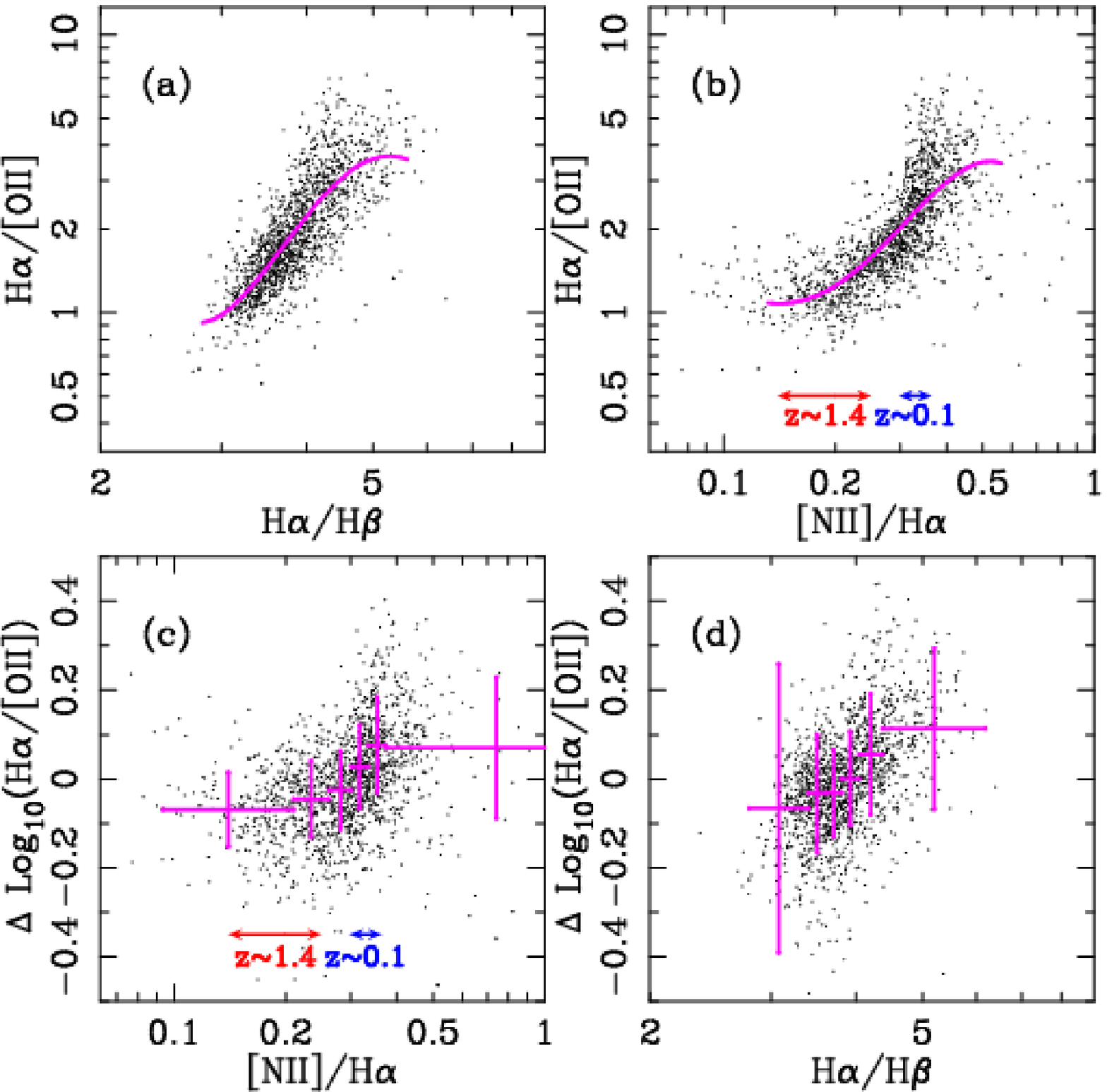} 
\end{center}
 \caption{ 
   (a) The \ha/\oii\ ratios for the SDSS sample as a function of the
   \ha/\hb\ ratio. The magenta line shows the polynomial function of the
   fourth degree which is best fitted to the data. 
   (b) Same as the upper left panel, but as a function of the
   \nii/\ha\ ratio. The red arrow shows the metallicity range inferred
   from the mass-metallicity relation of star-forming galaxies at $z\sim1.4$
   \citep{Yabe2012}, while the blue arrow shows that for star-forming
   galaxies at $z\sim0.1$ \citep{Tremonti2004}.
   (c) The offset of \ha/\oii\ ratio from the best fitted polynomial
   function shown in the upper left panel, as a function of \nii/\ha,
   i.e,~metallicity. The arrows are same ones as shown in the upper
   right panel. The magenta crosses shows the median values in each bin
   of \nii/\ha\ ratio.
   (d) Same as the lower left panel, but as a function of \ha/\hb,
   i.e,~dust extinction, where the best fitted function shown in the
   upper right panel is used.
}
 \label{fig;dustmetal}
\end{figure}

As discussed by \citet{Sobral2012a}, \ha/\oii\ ratios can be dependent
on both dust extinction and metallicity, because the two factors both 
influence the \oii\ luminosity more than the \ha\ luminosity. 
Therefore, before discussing the line ratios found in the previous
section (Figure \ref{fig;oiiha}), the SDSS sample with matched mass
distribution is used to evaluate how the two factors of
dust extinction and metallicity relate to the ratio of \ha/\oii\ and
to understand which factors are main driver of the difference in the line
ratio between \oii\ emitters at $z=1.47$ and galaxies at $z\sim0.1$. 
The Balmer decrement of \ha/\hb\ is an indicator of dust extinction,
while the ratio of \nii/\ha, the so-called N2 index, offers a rough
estimator of metallicity. The \ha/\oii\ ratios for the SDSS sample
described in \S\ref{sec;sdsssample} are plotted as a function of
\ha/\hb\ and \nii/\ha\ in Figure \ref{fig;dustmetal}, where a
polynomial function of the fourth degree is fitted to the data. The
figure implies that the ratio of \ha\ to \oii\ is dependent on both
factors in the sense that it increases with higher dust extinction and
higher metal abundance.

It is natural to expect that the \ha/\oii\ ratios are sensitive to dust,
since a large amount of dust prevents more \oii\ emissions at rest-frame
3727\AA\ from escaping from the star-forming regions in a galaxy than
\ha\ emissions at rest-frame 6563\AA. On the other hand, the
\ha/\oii\ line ratio is also dependent on the metallicity as shown in
Figure \ref{fig;dustmetal}. According to the metallicities estimated
from the mass-metallicity relations that recent studies have
suggested, the figure suggests that the difference in metallicity
between star-forming galaxies at $z\sim1.4$ and $z\sim0.1$ could
account for an offset of $\Delta\log$(\ha/\oii)$\sim0.3$dex. 

In reality, dust attenuation and metallicity are likely to be
correlated. To investigate the independent effect of metallicity on the
line ratio, Figure \ref{fig;dustmetal} also shows the offsets of
\ha/\oii\ ratio from the best fitted polynomial function shown in
Figure \ref{fig;dustmetal}, as a function of \nii/\ha, i.e,~metallicity. A
weak correlation is seen in the sense that the offset of
$\Delta\log$(\ha/\oii) gets larger with increasing \nii/\ha, although the
dispersion is large. At fixed dust attenuation, the difference in
metallicity between star-forming galaxies at $z\sim1.4$ and $z\sim0.1$
suggests an offset of at most $\sim0.1$dex in the mean value of
$\Delta\log$(\ha/\oii). This seems to suggest that the metallicity is
not the major factor on the offset of the \ha/\oii\ ratio. The same
suggestion is also obtained with larger sample of SDSS data by
\citet{Sobral2012a}. In contrast, Figure \ref{fig;dustmetal} shows the
offsets of \ha/\oii\ ratio from the best fitted polynomial function
shown in Figure \ref{fig;dustmetal}, as a function of \ha/\hb. This
figure suggests that the offset is larger with higher dust extinction
and the dependence of the offset on the dust extinction seems to be
stronger than the dependence seen in Figure \ref{fig;dustmetal} for
metallicity, although dispersion is still large. This may 
suggest that dust extinction is the more important factor in changing
the \ha/\oii\ ratio.

\subsection{Implications for dust extinction and metallicity}

\begin{figure}
\begin{center}
  \includegraphics[width=1.0\linewidth]{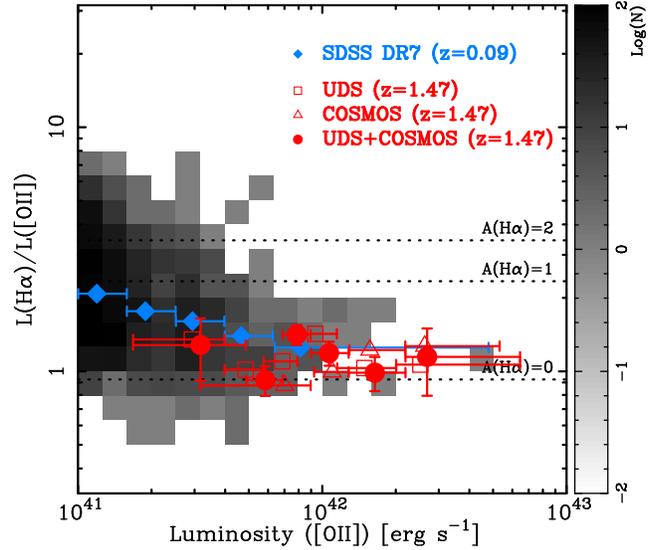} 
\end{center}
 \caption{ Same as Figure \ref{fig;oiiha}, but the ratio of \ha\ to
  \oii\ as a function of \oii\ luminosity. The flux ratios are
  corrected by 3\% for the bias caused by the difference of profile of
  the two narrow-band filters (\S \ref{sec;NBprofiles}) and 
  by 15\% for the  possible contamination to be included in the sample
  (\S \ref{sec;method}).  
  The dotted lines show the ratio of \ha\ to \oii\ in the case of each
  dust attenuation in \ha\ which is derived from the best-fitted
  polynomial function shown in Figure \ref{fig;dustmetal}.   
}
 \label{fig;oiiharatio}
\end{figure}

The average \ha/\oii\ ratios are plotted in Figure \ref{fig;oiiharatio}
as a function of \oii\ luminosity.
We find that they shift slightly towards a lower value at $2.8\sigma$
significance than those of galaxies at $z\sim0.1$.
According to the results of Figure \ref{fig;dustmetal}, the ratio
of \ha/\oii\ $\sim1$ implies that the typical \oii\ emitters at
$z=1.47$ are not dusty, metal-rich galaxies. Moreover, it is
unlikely that the metallicities of the \oii\ emitters at $z=1.47$
deviate significantly from the mass-metallicity relation that
\citet{Yabe2012} have found for galaxies at $z\sim1.4$.
That is, it seems that metallicity is not a main cause of the
difference in the \ha/\oii\ ratio, although we cannot completely rule
out the possibility that the observed line ratios are in part
dependent on the metallicity, as suggested by Figure \ref{fig;dustmetal}.  
We thus discuss the line ratios in terms of the dust extinction.

The lines in Figure \ref{fig;oiiharatio} show the \ha/\oii\
line ratio under the assumption that the ratios are changed due to the
amount of dust which is expressed as an attenuation in \ha, \Aha\
(the best-fitted polynomial function shown in Figure \ref{fig;dustmetal}).
In this case, all of the \ha/\oii\ ratios for the \oii\ emitters at
$z=1.47$ are distributed around the line corresponding to
\Aha$\sim$0.35, suggesting that \oii\ emitters at $z=1.47$ are likely
to be less subject to dust extinction. This result is not surprising
at all, since \oii\ emissions at rest-frame 3727\AA\ should be
sensitive to dust extinction.  
Although this is within expectations, it is important to directly
confirm with nebular emissions that typical \oii\ emitters are on
average a less dusty population over the wide range of the luminosity
down to $10^{41}$ erg~s$^{-1}$, i.e, SFR=1.4 \Msun yr$^{-1}$
\citep{kennicutt1998} at high redshift of $z=1.47$.  

Moreover, Figure \ref{fig;oiiharatio} suggests that local galaxies
selected based on \oii\ luminosity tend to be more subject to dust
attenuation compared with \oii\ emitters at $z=1.47$. However, the amounts
of dust attenuation are less than A(\ha)=1 (which is a typical value for
\ha\ emitters) in almost all bins of \oii\ luminosity.  On the other hand,
\citet{Sobral2012a} have found that \ha\ emitters at $z=1.47$ shows the
average \ha/\oii\ ratios consistent with those of the local galaxies with
A(\ha)$\sim$1. Thus, \oii\ emitters are likely to have smaller amount of
dust extinction on average than \ha\ emitters at both redshifts of
$z\sim0$ and 1.47, and more interestingly, typical \oii\ emitters at
$z=1.47$ seem to be less dusty than those in the local Universe.

Recent \oii\ emission surveys at $z\ga1$ have found that there are
\oii\ emitters with red colours comparable to red sequence on the
colour--magnitude diagram. However, the fact that \oii\ emitters at
$z\sim1.5$ typically show little dust attenuation implies that the red
\oii\ emitters are likely to be passive galaxies with AGN activity
in the core of galaxy, not dusty starburst galaxies. Indeed, we confirm
that mean \ha/\oii\ line ratios for \oii\ emitters with ($z-K$)$>$2.2 are
comparable to those for blue \oii\ emitters by conducting the stacking
analysis in the same manner as described above for the samples
classified by the colours. These results support the similar
conclusion on the population of red \oii\ emitters found in the galaxy
cluster $z=1.46$ \citep{hayashi2011} as well as those in lower
redshifts \citep{yan2006,lemaux2010,tanaka2012}. Moreover, the result
we have found in this paper highlights that surveys
of star formation activity based on only \oii\ emissions might result
in the underestimation to some extent, because \oii\ emission
surveys tend to be biased toward less dusty galaxies and therefore will
miss the most dusty starburst galaxies.

\section{Conclusions}
\label{sec;conclusions}

We investigate the mean relation between \ha\ and \oii\ luminosities
for \oii\ emitters at $z=1.47$ in the UDS and COSMOS fields using a
stacking analysis which enables us to examine the \ha\ luminosity of
galaxies at $z=1.47$ even if the individual galaxies are too faint to detect
both \ha\ and \oii\ emission lines simultaneously. The \oii\ emitters
at $z=1.47$ are selected with the NB921 narrow-band data taken with
Suprime-Cam on Subaru Telescope, while the \ha\ luminosities are
measured on the stacked \NBH\ narrow-band data taken with WFCAM on UKIRT.  

We find that on average there is positive correlation between \ha\ and
\oii\ luminosities for not only bright galaxies but also faint ones
with \oii\ luminosity down to $10^{41}$ erg s$^{-1}$, i.e, SFR=1.4
\Msun yr$^{-1}$. The trend that galaxies with higher \oii\
luminosities have higher \ha\ luminosities is consistent with that of
the local galaxies, suggesting that \oii\ luminosities can be used as
an indicator of SFR even at the high redshift of $z=1.47$.  

However, we have to use the \oii\ luminosities with caution to
estimate SFRs at $z=1.47$ based on the relation calibrated with local
galaxies.
This is because \oii\ emitters at $z=1.47$ show observed \ha/\oii\
line ratios corresponding to \Aha$\sim$0.35 and are less subject to
dust attenuation than the local galaxies selected based on \oii\
luminosity. 
Therefore, \oii -selected emitters at $z=1.47$ are biased toward less
dusty populations.  
The use of dust-correction relations derived with \ha\ emitter samples
may cause us to overestimate the amount of dust extinction and hence
the dust-corrected SFR. At the same time, surveys of star formation
activity based on \oii\ emissions may miss populations of dusty
starburst galaxies.  

On the other hand, we note a caveat to our interpretation of the
results in terms of dust extinction only, because the
\ha/\oii\ line ratio is also dependent on the metallicity. Hence the
discrepancy of the line ratio between \oii\ emitters at $z=1.47$ and
local galaxies $z\sim0.1$ may be explained in terms of metallicity
difference. Therefore, the possibility that the low \ha/\oii\ ratio is
not only due to the lower dust extinction, but also the lower
metallicities of \oii\ emitters at $z=1.47$ than local galaxies,
cannot be completely ruled out. To distinguish the two factors of dust
extinction and metallicity completely, we require deep spectroscopy to
obtain the nebular emissions from the individual or stacked spectra.    

\section*{Acknowledgments}
We would like to thank an anonymous referee for carefully reading our
manuscript and providing helpful comments. 
MH is grateful for financial support from the Japan Society for the
Promotion of Science (JSPS) fund, ``Institutional Program for  
Young Researcher Overseas Visits'' to stay at the IfA, Royal
Observatory of Edinburgh for two months in 2011.   
DS acknowledges financial support from the Netherlands Organisation
for Scientific research (NWO) through a Veni fellowship and the NOVA
Research school for a NOVA fellowship. 
IRS acknowledges support from the Leverhulme Trust and STFC. 
The United Kingdom Infrared Telescope is operated by the Joint
Astronomy Centre on behalf of the Science and Technology Facilities
Council of the U.K. Data used in this paper are in part collected at
Subaru Telescope, which is operated by the National Astronomical
Observatory of Japan.

\bsp

\label{lastpage}

\end{document}